\documentclass{phostproc}

\usepackage[urlcolor=magenta]{hyperref}
\usepackage{subfigure}

\title{Synergy between asteroseismology and exoplanet science:\\ an outlook}
\author{Tiago L.~Campante$^{1,2}$, Susana C.~C.~Barros$^{1,2}$, Olivier Demangeon$^{1}$, Hugo J.~da N\'obrega$^{2,3}$, \\ James S.~Kuszlewicz$^{4,5}$, Filipe Pereira$^{1}$, William J.~Chaplin$^{6,5}$, and Daniel Huber$^{7}$}

\affiliation{
$^{1}$ Instituto de Astrof\'{\i}sica e Ci\^{e}ncias do Espa\c{c}o, Universidade do Porto, Rua das Estrelas, PT4150-762 Porto, Portugal \\
Email: \href{mailto:tiago.campante@astro.up.pt}{Tiago.Campante@astro.up.pt} \\
$^{2}$ Departamento de F\'{\i}sica e Astronomia, Faculdade de Ci\^{e}ncias da Universidade do Porto, Rua do Campo Alegre, s/n, PT4169-007 Porto, Portugal \\
$^{3}$ CAP -- Centre for Applied Photonics, INESC TEC, Porto, Portugal \\
$^{4}$ Max Planck Institute for Solar System Research, D-37077 G\"ottingen, Germany \\
$^{5}$ Stellar Astrophysics Centre (SAC), Department of Physics and Astronomy, Aarhus University, Ny Munkegade 120, DK-8000 Aarhus C, Denmark \\
$^{6}$ School of Physics and Astronomy, University of Birmingham, Birmingham B15 2TT, UK \\
$^{7}$ Institute for Astronomy, University of Hawai`i, 2680 Woodlawn Drive, Honolulu, HI 96822, USA
}

\shorttitle{Asteroseismology and exoplanet science}
\shortauthors{Tiago L.~Campante \textit{et al.}}

\abs{Space-based asteroseismology has been playing an important role in the characterization of exoplanet-host stars and their planetary systems. The future looks even brighter, with space missions such as NASA's TESS and ESA's PLATO ready to take on this legacy. In this contribution, we provide an outlook on the synergy between asteroseismology and exoplanet science, namely, on the prospect of conducting a populational study of giant planets around oscillating evolved stars with the TESS mission.}

\begin{document}

\maketitle

\section{Introduction}
The asteroseismology revolution initiated by \textit{Kepler} \citep{Kepler} is set to continue over the coming decades with the launches of TESS \citep{TESS}, PLATO \citep{PLATO}, as well as WFIRST \citep{WFIRST}, with these missions expected to raise the number of solar-like oscillators to a few million stars \citep{Huber18}. Note that over 90\,\% of all detections are expected to be for evolved stars, with PLATO by far contributing the most detections for dwarfs and subgiants ($\sim\!80{,}000$). If we combine this with dedicated ground-based efforts, such as the SONG network \citep{SONG} of 1-meter telescopes, we are then positive that the synergy between asteroseismology and exoplanet science can only continue to grow \citep{CampanteBook}.

Synergetic studies of evolved stars are made possible by even moderate photometric cadences, which can be used to simultaneously detect transits and stellar oscillations. A very exciting prospect is that of conducting asteroseismology of red-giant hosts using the 30-minute cadence of TESS full-frame images (FFIs). Based on an all-sky stellar and planetary synthetic population \citep{Sullivan15}, we predict that solar-like oscillations will be detectable in up to 200 low-luminosity red-giant branch (LLRGB) stars hosting close-in giant planets \citep{Campante16}.

The population of transiting planets around evolved stars is so far largely unexplored \citep{Huber18}. And although radial-velocity surveys are mostly complete for planets near or above $1\,M_{\rm Jup}$ at $> 0.2\:{\rm AU}$, there is a dearth of planets with orbital periods $P < 80\:{\rm d}$. Nonetheless, \textit{Kepler}/K2 have discovered several close-in giant planets around LLRGB stars \citep[e.g.,][]{Grunblatt16,Grunblatt17}, hinting at a population of warm sub-Jovian planets around evolved stars that would be accessible to TESS. \textit{Kepler}/K2 mainly targeted main-sequence stars, and observed too few LLRGB stars to detect enough planets for robust statistics. TESS will increase the number of LLRGB stars with space-based photometry by one order of magnitude over \textit{Kepler}/K2, providing an unprecedented opportunity to address a number of key questions in exoplanet science, namely:
\begin{itemize}
\item The role of stellar flux on hot-Jupiter inflation;
\item Giant-planet occurrence as a function of stellar mass and evolution;
\item Correlation between metallicity and giant-planet occurrence around evolved stars.
\end{itemize}

In this work, we start by characterizing the parent population of LLRGB stars to be searched for transits based on a {\sc galaxia} \citep{Sharma11} simulation (Sect.~\ref{sec:parent}). We focus on the southern ecliptic hemisphere, which will be surveyed during year 1 of TESS's primary mission. We next implement a software tool for planetary-transit search based on a Box-fitting Least Squares (BLS) algorithm (Sect.~\ref{sec:detect}). The tool is tested both for statistical false positive rates and detection sensitivity using artificial TESS light curves. The former involves running the code on a sufficiently large number of light curves containing only instrumental/shot noise and stellar (correlated) signals, namely, granulation and oscillations. The latter involves running the code on the same light curves, although now with injected transits. We conclude with a few considerations regarding the use of TESS photometry alone in candidate vetting (Sect.~\ref{sec:vetting}).

\section{Parent stellar population}\label{sec:parent}

\subsection{Synthetic population}\label{sec:synth}
We start with an all-sky, magnitude-limited synthetic stellar population generated with {\sc galaxia}. Output absolute magnitudes were converted to apparent magnitudes and extinction applied. We ended up only retaining stars down to magnitude 13 in the Johnson--Cousins $I_{\rm C}$ band. Although somewhat optimistic, this magnitude cut is used to ensure that all detectable oscillating LLRGB stars are captured \citep{Campante16}. Note that the simulation is undersampled by a factor of 10 to ease up on the data handling.

We made an initial selection of \textit{putative} LLRGB stars from this synthetic population by applying the following cuts on $\log g$ and $T_{\rm eff}$: $2.7 < \log g < 3.5$ and $T_{\rm eff} < 5500\,{\rm K}$. The upper $\log g$ cut ensures that stars are evolved enough to oscillate with frequencies detectable with 30-minute-cadence data \citep{Chaplin14}. The lower $\log g$ cut is purely empirical \citep{Hekker11} and leads to contamination of the sample by red clump (RC) stars (on which more below). This step was followed by selecting only those stars located in the southern ecliptic hemisphere. Finally, we determined the median number of sectors (by considering a number of different initial pointings) over which each star would be observed with TESS using the {\sc Python} package {\sc tvguide}\footnote{\url{https://heasarc.gsfc.nasa.gov/docs/tess/proposal-tools.html\#tvguide}} (each sector corresponds to a 27.4-day coverage), having discarded stars that fall off silicon. The final tally amounts to $\sim6.8\times10^5$ stars (after applying the factor-of-10 correction).

\begin{figure}[!t]
\centering
  \includegraphics[width=\linewidth]{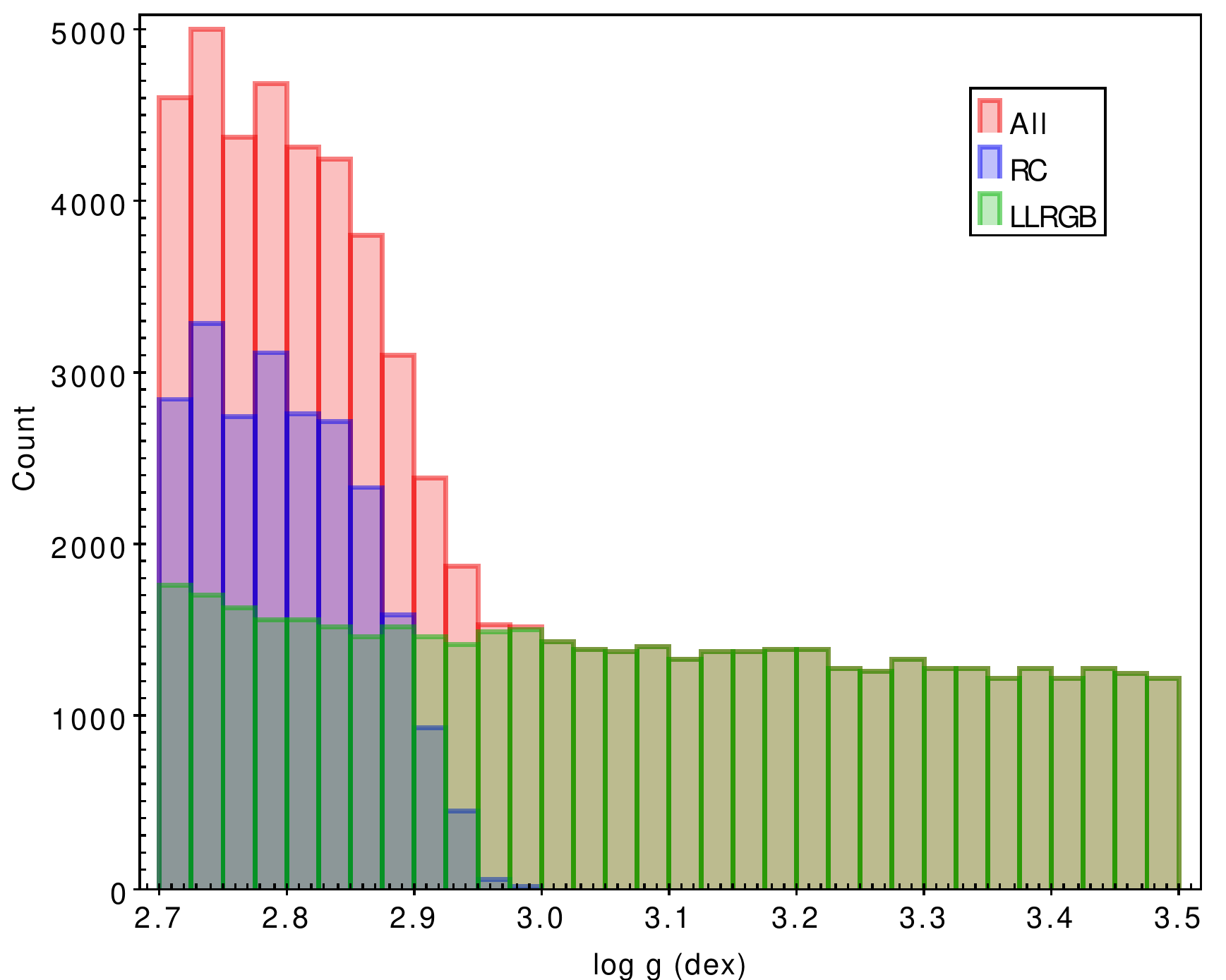}
  \caption{\small Contamination of RC stars in the initial sample.}\label{fig:RC_vs_LLRGB}
\end{figure}

We now come back to the issue of the sample contamination by RC stars. Indeed, 34\,\% of stars in the above sample are RC stars (see Fig.~\ref{fig:RC_vs_LLRGB}). We apply an additional cut on the asteroseismic observable $\Delta\nu$, namely, $\Delta\nu > 10\:{\rm \mu Hz}$, in order to mitigate this contamination effect. This effectively leads to the removal of all RC stars and is the equivalent to raising the lower $\log g$ cut to $\log g \gtrsim 2.9$. We take the resulting sample of bona fide LLRGB stars as our final, parent (synthetic) stellar population, which comprises $\sim3.0\times10^5$ stars (after applying the factor-of-10 correction).

\subsection{Population characteristics}
\begin{figure}[!t]
\centering
  \subfigure[]{%
  \includegraphics[width=\linewidth]{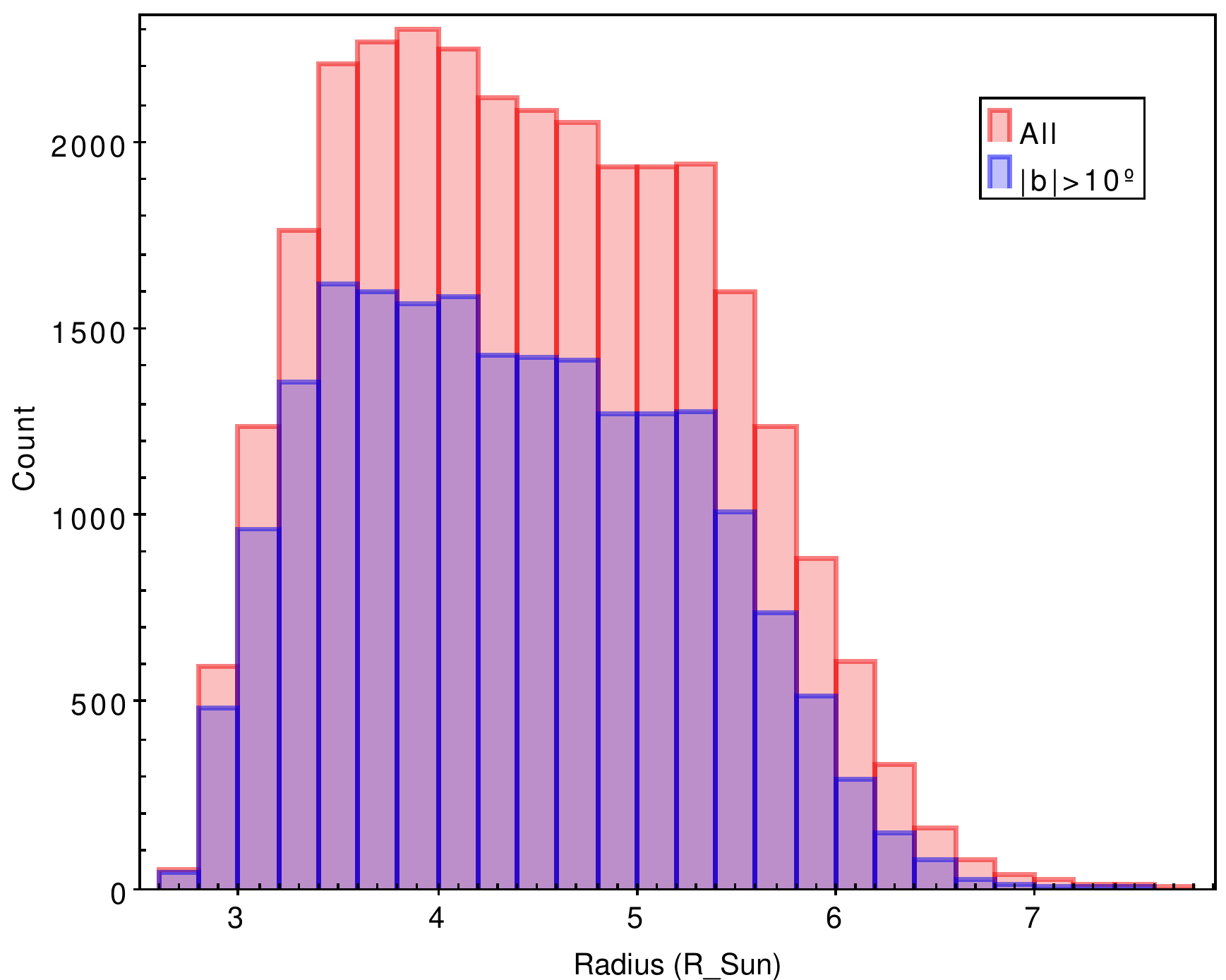}}
  \subfigure[]{%
  \includegraphics[width=\linewidth]{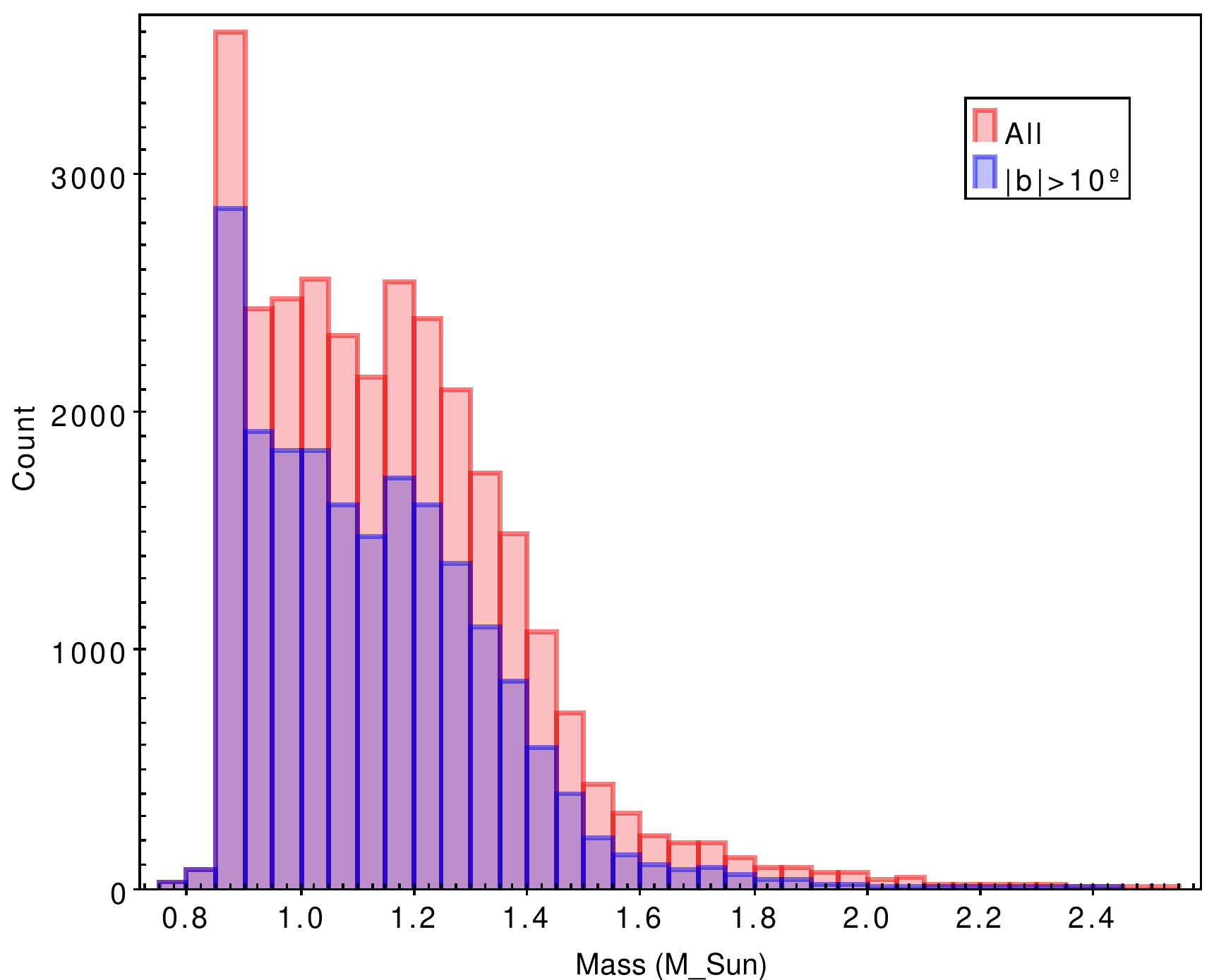}}\\
  \subfigure[]{%
  \includegraphics[width=\linewidth]{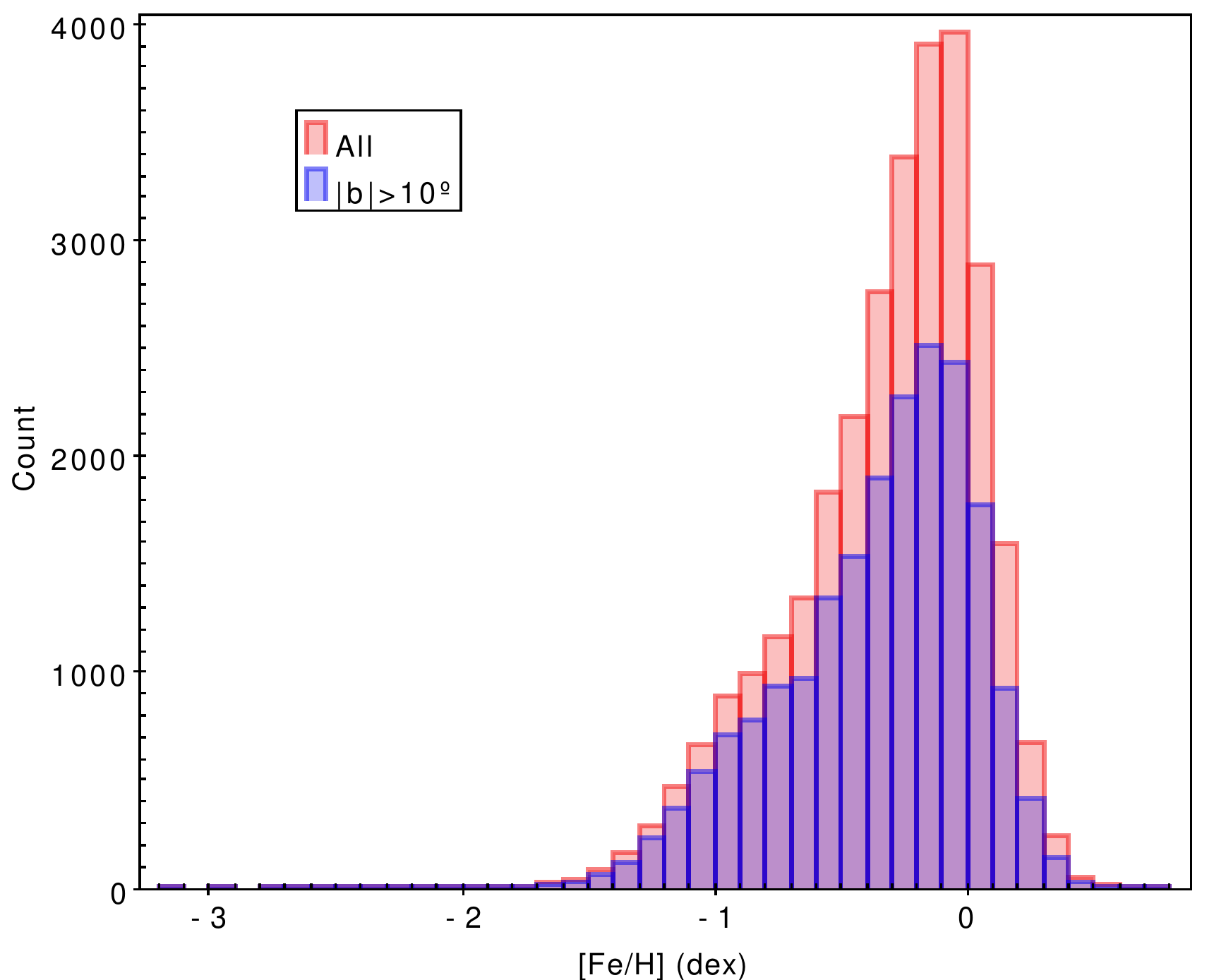}}
  \caption{\small Stellar (a) radius, (b) mass and (c) metallicity distributions of the parent stellar population.}\label{fig:properties}
\end{figure}

We now look at the main characteristics of this parent (synthetic) stellar population. Figure \ref{fig:properties} shows the overall stellar radius, mass and metallicity distributions (light red). Distributions are also shown for a subset of stars lying farther away from the Galactic plane, i.e., $|b|>10^\circ$ (light blue; comprising 68\,\% of the parent population). The Galactic latitude of a target strongly influences the likelihood of it giving rise to an astrophysical false positive. For $|b|<10^\circ$, the density of background stars is very high, meaning that any observed eclipse is more likely to be from a background eclipsing binary, whereas for $|b|>20^\circ$ planets should represent a majority over false positives\footnote{Note that these remarks were made with reference to planet detections around TESS target stars.} \citep{Sullivan15}. 

\begin{figure}[!t]
\centering
  \subfigure[]{%
  \includegraphics[width=\linewidth]{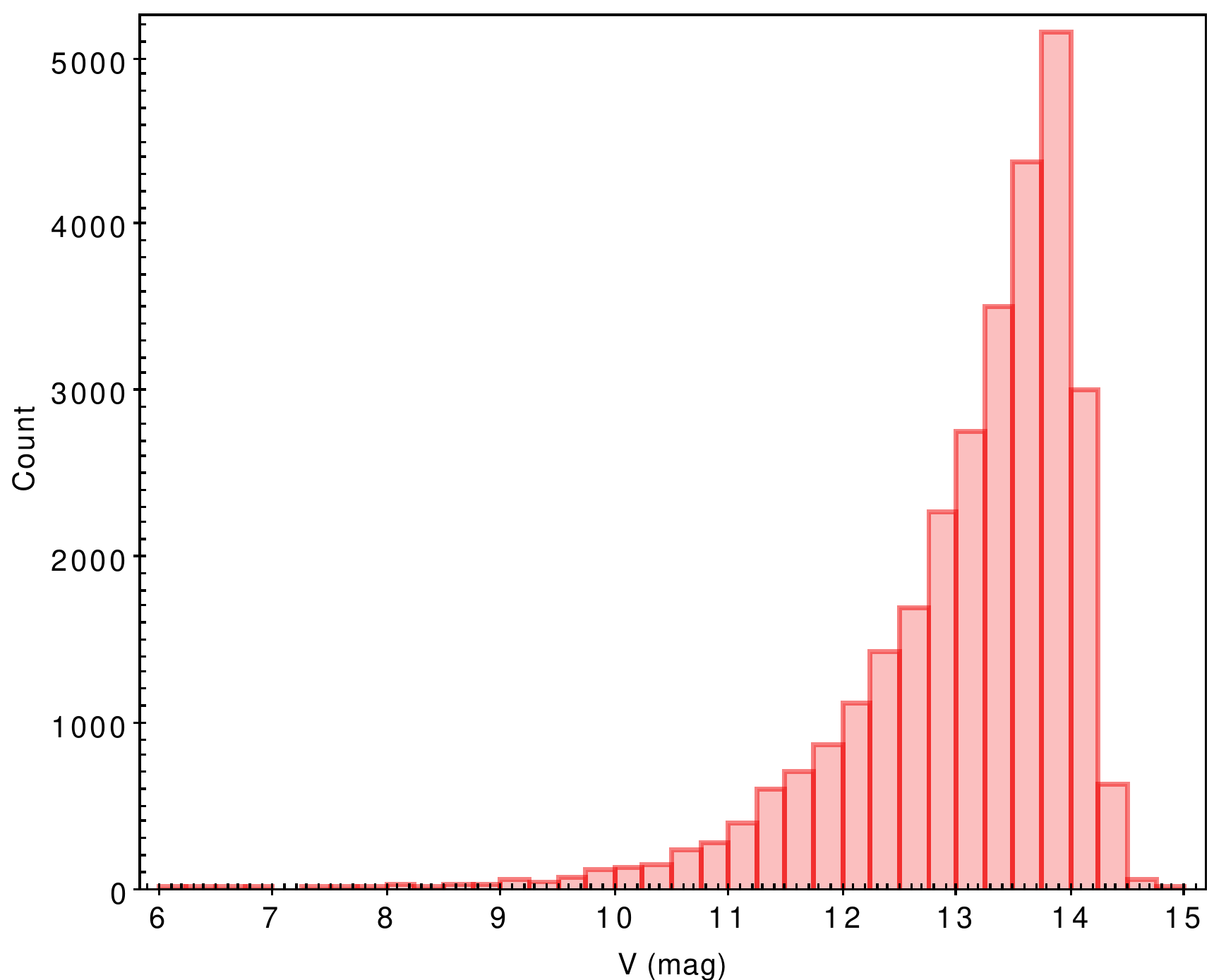}}
  \subfigure[]{%
  \includegraphics[width=\linewidth]{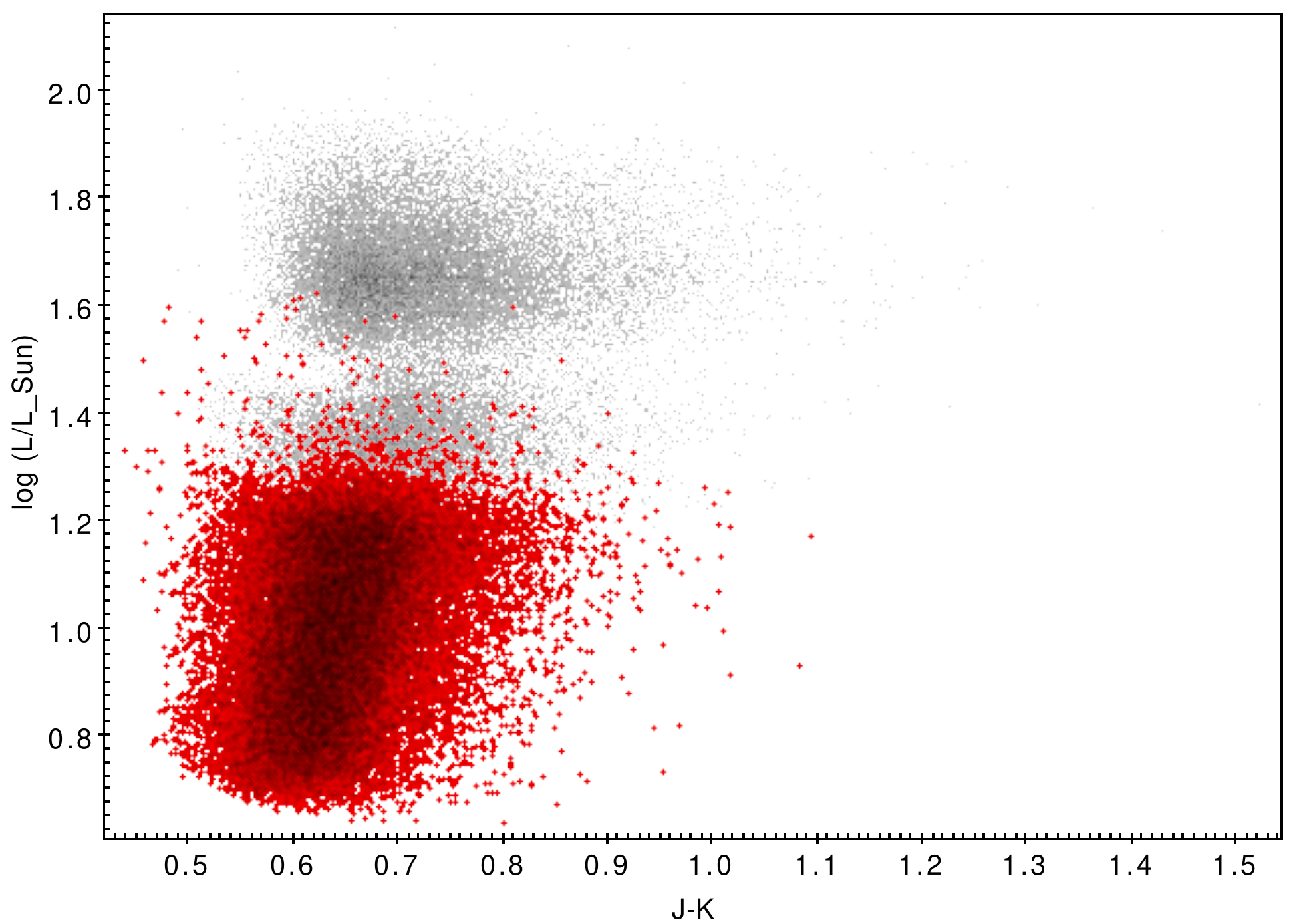}}
  \caption{\small $V$-band magnitude distribution (a) and luminosity-color diagram (b) of the parent stellar population. The luminosity-color diagram of the parent stellar population (red) is superimposed on that of the initial sample (gray; see Sect.~\ref{sec:synth}).}\label{fig:properties2}
\end{figure}

Panel (a) of Fig.~\ref{fig:properties2} shows the $V$-band magnitude distribution of the parent population (peaking at $V \sim 13$--14), while panel (b) shows a luminosity-color diagram that may be used to inform target selection. We will likely need to apply a stricter magnitude cut depending on the actual oscillations detectability limit. We also assessed the fraction of stars in the parent population observed over a median of 1 TESS sector (78\,\%) and 1 or 2 TESS sectors (93\,\%). Furthermore, we note that for 26\,\% of the stars $\nu_{\rm max}$ is above the Nyquist frequency.

\section{Automated transit detection: performance assessment}\label{sec:detect}

\subsection{Artificial light curves}\label{sec:art_lc}
We generated artificial light curves for the $\sim\!30{,}000$ unique, bona fide LLRGB stars in our parent (synthetic) stellar population (see Sect.~\ref{sec:parent}). Generation of the light curves is performed originally in the frequency domain, after which an inverse Fourier transform is applied (Kuszlewicz et al., submitted). We consider only the 30-minute cadence of TESS FFIs and apply a window function to account for the data downlink occurring every spacecraft orbit.

\begin{figure}[!h]
\centering
  \subfigure[]{%
  \includegraphics[width=\linewidth]{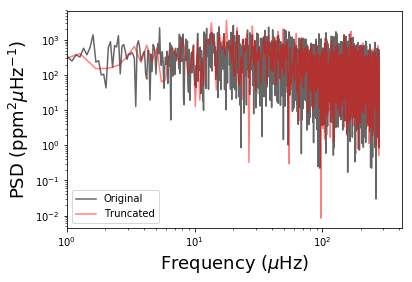}}
  \subfigure[]{%
  \includegraphics[width=\linewidth]{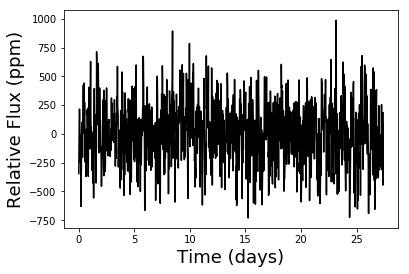}}
  \caption{\small Power spectral density (a) and corresponding light curve (b) of a $V=10.3$ star observed for 27.4 days (or 1 TESS sector). The star has $\nu_{\rm max} \approx 167\:{\rm \mu Hz}$. The resolution of the truncated spectrum (red) is defined by the duration of the light curve, whereas the original spectrum (black) is oversampled. No window function has yet been applied to the light curve to account for the data downlink.}\label{fig:example_ps_ts}
\end{figure}

We used a photometric noise model for TESS \citep{Sullivan15,Campante16} to predict the rms noise per exposure time. A systematic term of $20\:{\rm ppm\,hr^{1/2}}$ was included in this calculation. To model the granulation power spectral density, we adopted a scaled version (to predict TESS granulation amplitudes) of model F of \citet{Kallinger14}, which contains two Harvey-like components. No aliased granulation power was considered. Individual radial, (mixed) dipole and quadrupole modes were also modeled whenever $\nu_{\rm max} \! < \! \nu_{\rm Nyq}$. Panels (a) and (b) of Fig.~\ref{fig:example_ps_ts} respectively display the power spectral density (PSD) and corresponding light curve of a $V=10.3$ star observed for 27.4 days (or 1 TESS sector). The oscillation bump can be seen around $\nu_{\rm max} \! \approx \! 167\:{\rm \mu Hz}$.

We generated model transit light curves using the {\sc Python} package {\sc batman}\footnote{\url{https://www.cfa.harvard.edu/~lkreidberg/batman/}}. Assuming circular planetary orbits, we seeded one planet per star. We next drew orbital periods and planet radii from uniform distributions spanning the parameter space of interest (0.5 to 27.4 days and 4 to 22 $R_\oplus$, respectively). Orbital periods were redrawn until no systems were left within the Roche limit or the stellar envelope. We assume that all planets transit and draw the impact parameter from a uniform distribution defined over the half-open interval $[0,1[$. Input to {\sc batman} includes the time of inferior conjunction, orbital period, planet radius, semi-major axis, and orbital inclination. A quadratic limb darkening law is used and its coefficients set to fixed values \citep[see][]{Barclay15}. We further account for the long integration time by supersampling the model 11 times per cadence then integrating over these subsamples.

\subsection{BLS algorithm}\label{sec:bls}
We search for transits using an updated version of the pipeline presented in \citet{Barros16}, which makes use of a {\sc Python} implementation\footnote{\url{https://github.com/dfm/python-bls}} of the BLS algorithm originally introduced by \citet{Kovacs02}. The search is made over periods ranging from 1 day to 70\,\% of the light curve duration and over fractional transit durations ranging from 0.001 to 0.3 with $nb=200$ phase bins. Frequency sampling is optimized according to $\delta\nu=1/(P_{\rm max} \cdot nb)$, where $P_{\rm max}$ is the maximum period searched for. Using the periods and epochs found by the BLS algorithm, each light curve is phase-folded and the signal detection efficiency \citep[\textit{SDE};][]{Kovacs02} computed.

\begin{figure*}[!t]
\centering
  \includegraphics[width=\linewidth]{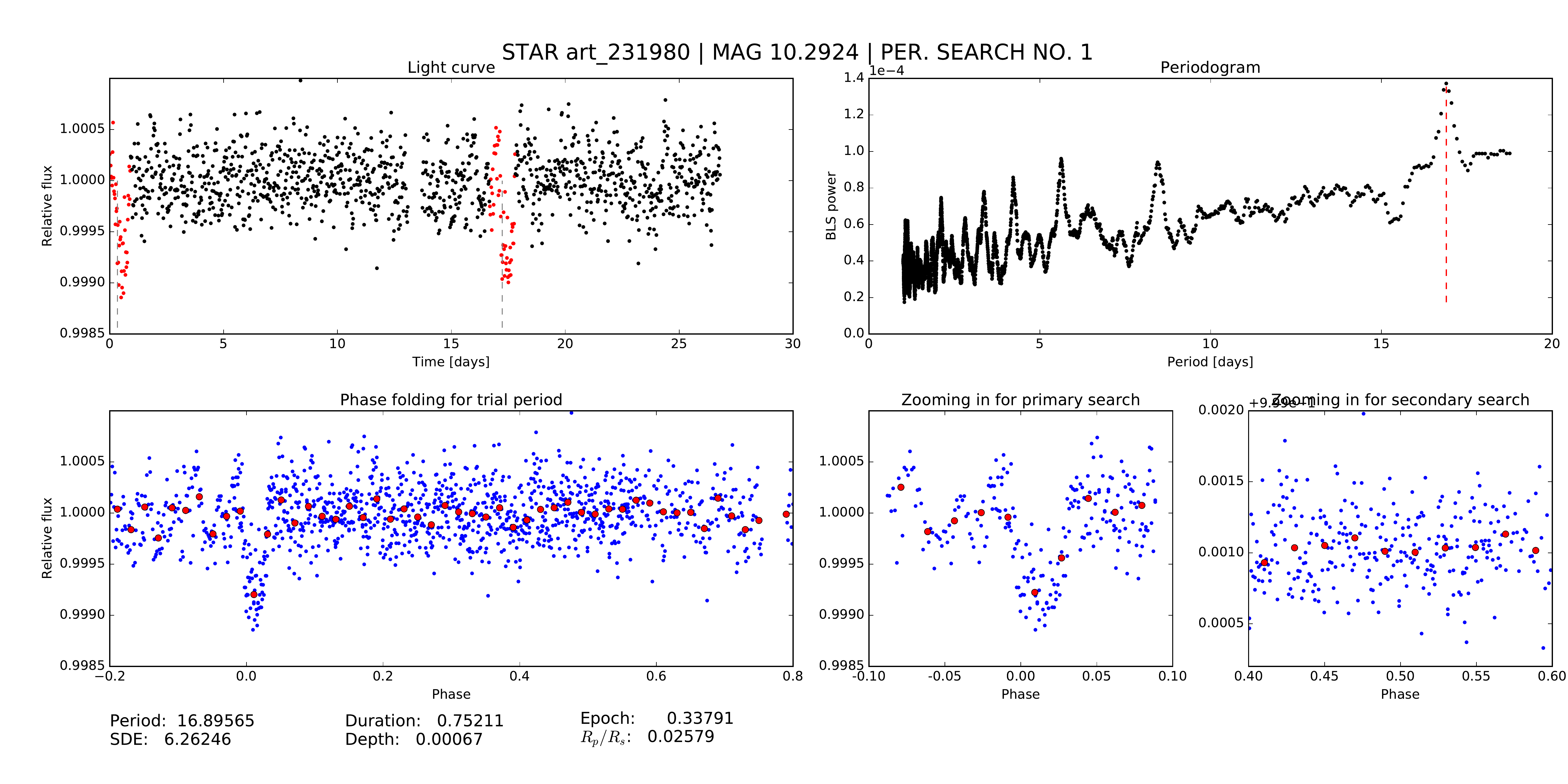}
  \caption{\small Pipeline output for the same artificial star considered in Fig.~\ref{fig:example_ps_ts}. The light curve is shown in the top left panel with both (correctly) recovered transits in red. Notice the gap at $\approx 13.7\:{\rm d}$ due to the data downlink. The BLS periodogram is shown in the top right panel with the vertical dashed line indicating the best period, as determined by the algorithm. The bottom left panel displays the phase-folded light curve using the best period (blue) and a binned version of it (red). The bottom middle and right panels simply zoom in on the phase-folded light curve at the locations of the primary and possible secondary, respectively.}\label{fig:bls_output}
\end{figure*}

\begin{figure*}[!t]
\centering
  \includegraphics[width=\linewidth]{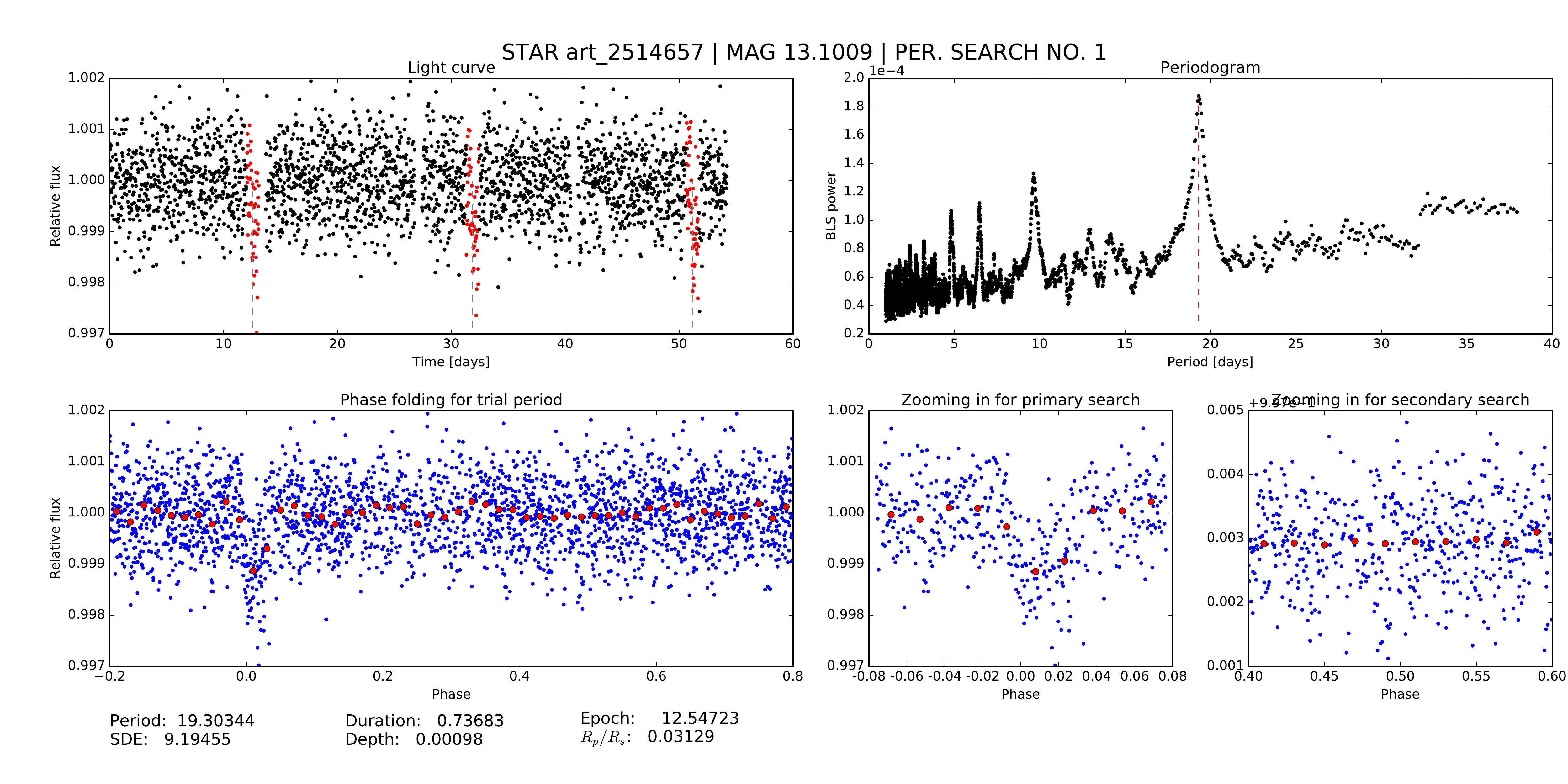}
  \caption{\small Pipeline output for a $V=13.1$ star observed for 54.8 days (or 2 TESS sectors). Panels are the same as in Fig.~\ref{fig:bls_output}.}\label{fig:bls_output2}
\end{figure*}

The pipeline searches $npass$ transits per light curve and sorts them according to the \textit{SDE}. Results for all the light curves are also sorted according to the maximum \textit{SDE} reported for each light curve. It also tests for the following features: possibility of a secondary transit/eclipse, sinusoidal behavior, and mono transit (or an effective number of transits that is less than the total number of transits). Provided the fitted depth is positive, the pipeline produces a series of plots for each candidate. Figure \ref{fig:bls_output} shows the pipeline output for the same artificial star considered in Fig.~\ref{fig:example_ps_ts} (after transit injection). The two injected transits are correctly recovered. Another example can be found in Fig.~\ref{fig:bls_output2}, where the pipeline output is shown for a $V=13.1$ star observed for 54.8 days (or 2 TESS sectors). All three injected transits are correctly recovered.

\subsection{Statistical false positives}\label{sec:fp}
\begin{figure}[]
\centering
  \subfigure[]{%
  \includegraphics[width=\linewidth]{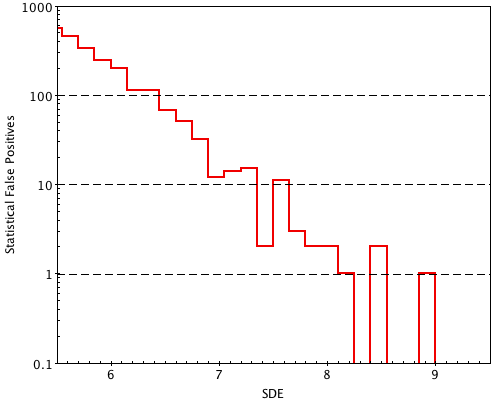}}\\
  \subfigure[]{%
  \includegraphics[width=\linewidth]{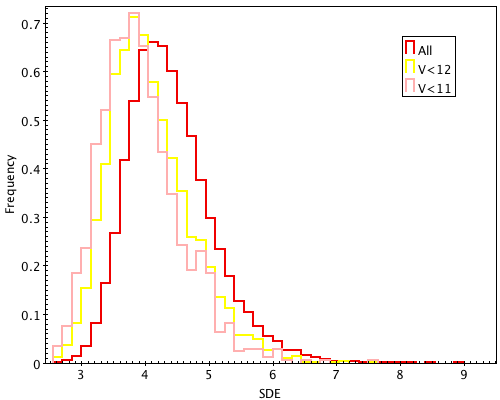}}
  \subfigure[]{%
  \includegraphics[width=\linewidth]{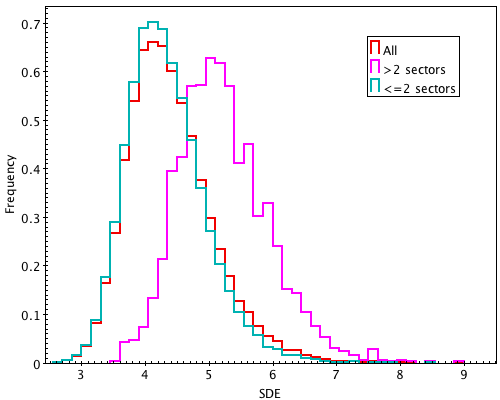}}
  \caption{\small Statistical false positive rates. Panel (a): Number of statistical false positives as a function of the \textit{SDE}. Horizontal dashed lines indicate 1, 10, and 100 false positives. Panels (b) and (c): Normalized histograms of the \textit{SDE} depicting its dependence on limiting magnitude and light curve duration, respectively.}\label{fig:threshold}
\end{figure}

We began by running the pipeline on the $\sim30{,}000$ generated artificial light curves (prior to transit injection) in order to assess the rate of statistical false positives. Results are shown in Fig.~\ref{fig:threshold}. From panel (a), we find that an \textit{SDE} threshold of 8.88 produces approximately one statistical false positive over the $\sim30{,}000$ light curves or a rate of 0.003\,\%. Table \ref{table:threshold} provides \textit{SDE} thresholds as a function of the statistical false positive rate (0.1\,\%, 1\,\% and 5\,\%). Panels (b) and (c) of Fig.~\ref{fig:threshold} emphasize the dependence of the \textit{SDE} on the sample's limiting magnitude and light curve duration, respectively. The dependence on the latter is particularly obvious, with shorter light curves giving rise to lower \textit{SDE} values \citep[cf.][]{Kovacs02}. Therefore, Table \ref{table:threshold} also provides \textit{SDE} thresholds when only 27.4- (1 sector) and 54.8-day-long (2 sectors) light curves are considered.

\begin{table}[!t]
\centering
\caption{\textit{SDE} threshold as a function of the statistical false positive rate.}
\label{table:threshold}
\begin{tabular}{cccc}
\hline\hline
Rate & \textit{SDE} (All) & \textit{SDE} (1 sector) & \textit{SDE} (2 sectors) \\
\hline
0.1\,\% & 7.29 & 6.75 & 7.22 \\
1\,\% & 6.34 & 5.98 & 6.39 \\
5\,\% & 5.61 & 5.33 & 5.73 \\
\hline
\end{tabular}
\end{table}

\subsection{Detection sensitivity}\label{sec:sensitivity}
\begin{figure}[!t]
\centering
  \includegraphics[width=\linewidth]{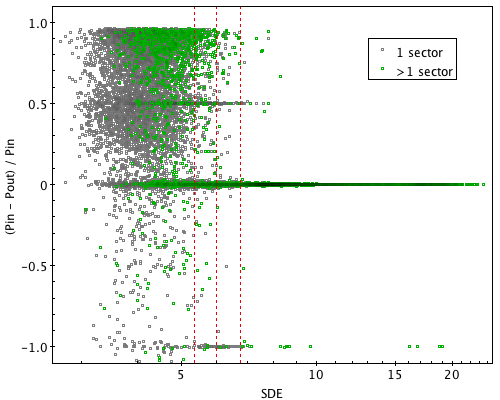}
  \caption{\small Relative difference between injected ($P_{\rm in}$) and recovered ($P_{\rm out}$) orbital periods as a function of the \textit{SDE}. Only systems with two or more detected transits and measured positive depths have been displayed. Vertical dashed lines mark (from right to left) the \textit{SDE} thresholds corresponding to 0.1\,\%, 1\,\% and 5\,\% statistical false positive rates when considering 1 TESS sector (see Table \ref{table:threshold}). Systems observed over 1 (gray) or more (green) TESS sectors are identified.}\label{fig:detections}
\end{figure}

We now run the pipeline on the same $\sim30{,}000$ artificial light curves after transit injection in order to assess the detection sensitivity (or survey completeness). Figure \ref{fig:detections} shows the relative difference between the injected ($P_{\rm in}$) and recovered ($P_{\rm out}$) orbital periods as a function of the \textit{SDE}. Vertical dashed lines mark (from right to left) the \textit{SDE} thresholds corresponding to 0.1\,\%, 1\,\% and 5\,\% statistical false positive rates when considering 1 TESS sector. 

The pipeline sometimes recovers half ($0.5$ ordinate) or double ($-1$ ordinate) the injected period, which tends to happen close to the aforementioned thresholds. The former subset mostly corresponds to the case of two injected transits with an orbital period exceeding 70\,\% of the light curve duration, whereas the latter, less numerous subset mostly corresponds to those injected transits with periods shorter than 1 day. These injected periods are out of bounds with respect to the search parameters of the algorithm. The remaining, rarer cases seem to be genuine statistical misidentifications prompted by the low \textit{SDE}.

\begin{figure}[!t]
\centering
  \includegraphics[width=\linewidth]{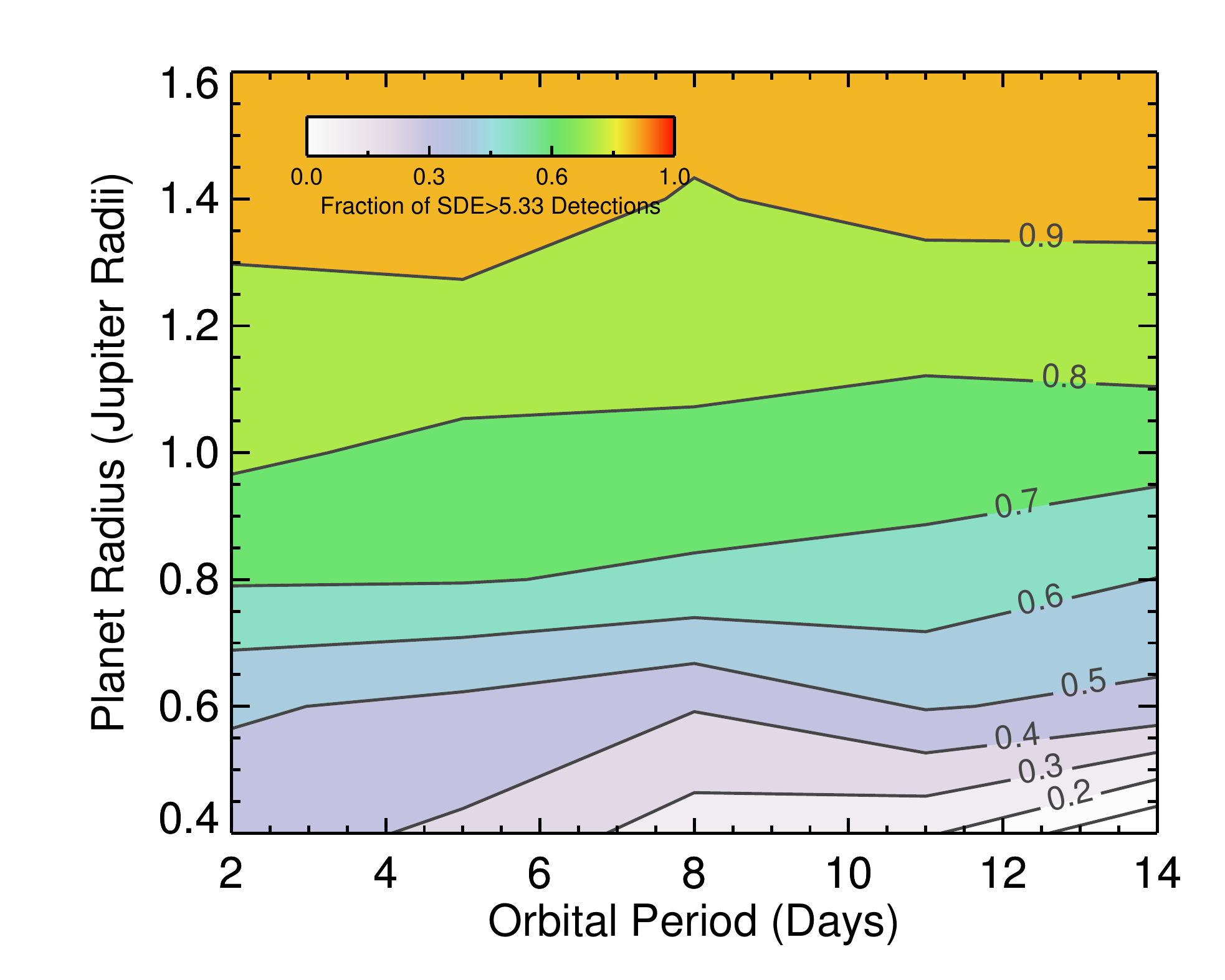}
  \caption{\small Transit detection sensitivity based on 27.4 days of data and a minimum of two transit events. We assume a statistical false positive rate of 5\,\% (or an \textit{SDE} threshold of 5.33; see Table \ref{table:threshold}).}\label{fig:sensitivity}
\end{figure}

Figure \ref{fig:sensitivity} shows the transit detection sensitivity based on 27.4 days of data and a minimum of two transit events, where we have assumed a statistical false positive rate of 5\,\%. The resulting contours demonstrate that even for the most pessimistic case of 27.4 days coverage it will be possible to detect close-in, inflated Jupiters for over 80\,\% of stars, Jupiter-size planets for 70--80\,\% of stars, and large Neptunes only in the most favorable cases.

\subsubsection{Out-of-transit flux modulation}
\begin{figure}[!t]
\centering
  \subfigure[]{%
  \includegraphics[width=\linewidth]{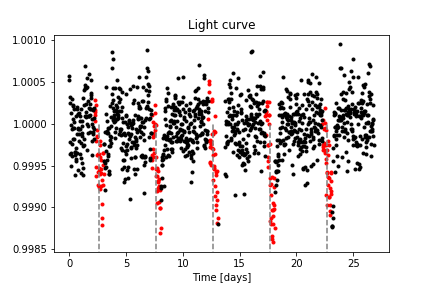}}
  \subfigure[]{%
  \includegraphics[width=\linewidth]{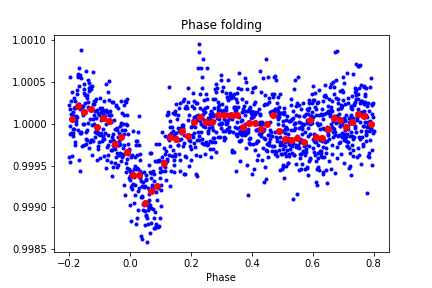}}
  \caption{\small Light curve (a) and phase-folded light curve (b) of an artificial system resembling the Kepler-91 system. The star is observed for 27.4 days (or 1 TESS sector). A sine function with period $P_{\rm in}/2$ and phase 0 at minimum has been injected into the light curve. Its amplitude was set to $100\:{\rm ppm}$. An ellipsoidal modulation is clearly seen in the right panel.}\label{fig:ellipsoidal}
\end{figure}

We test for the presence of sinusoidal behavior in the light curve by first fitting a sine function to the binned, folded light curve (on the best period). We then perform a linear regression (after linearization) to obtain the coefficient of determination, $r^2$. Here, we apply this procedure to an artificial system resembling the confirmed Kepler-91 planetary system, for which ellipsoidal variations have been measured that are caused by a close-in giant planet \citep{Box14,Barclay15}. We injected a sine function with period $P_{\rm in}/2$ and phase 0 at minimum into the light curve \citep{Pfahl08}, having varied its amplitude. The star is observed for 27.4 days only (or 1 TESS sector). A notional amplitude of $100\:{\rm ppm}$ of the ellipsoidal modulation (see Fig.~\ref{fig:ellipsoidal}) leads to $r^2=0.22$, above the $r^2=0.1$ threshold adopted within the pipeline to flag potential sinusoidal behavior. This amplitude is slightly higher than the approximate upper limit of $75\:{\rm ppm}$ measured for Kepler-91. We conclude by noting that it is usually necessary to remove instrumental trends and rotational modulation due to spots/plages prior to any transit search. Such detrending of the light curve may (depending on the period) also remove part of the out-of-transit flux modulation, thus making detection of ellipsoidal variations more challenging.

\subsubsection{Secondary transit/eclipse}
\begin{figure*}[!t]
\centering
  \subfigure[]{%
  \includegraphics[width=\textwidth]{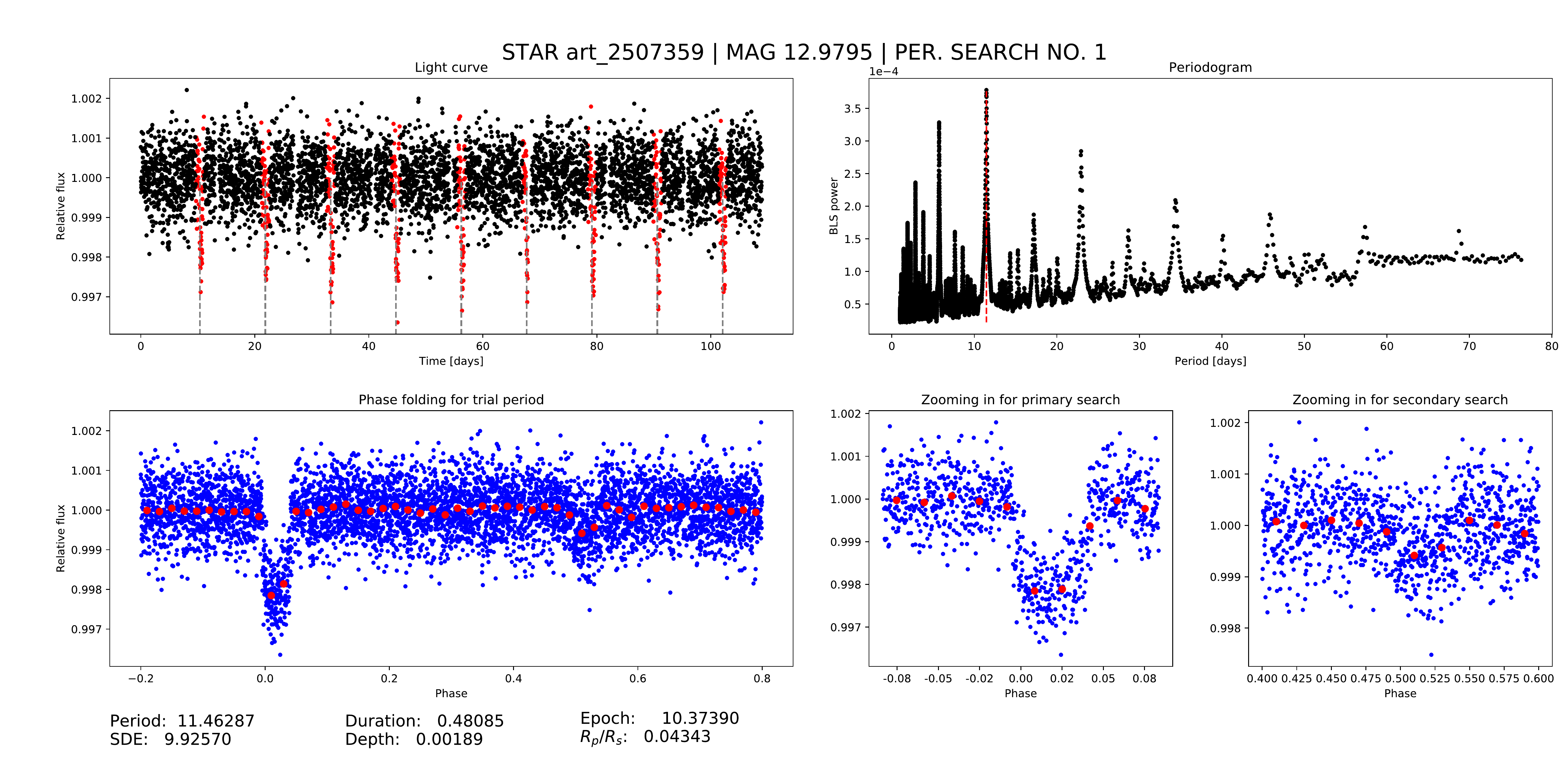}}
  \subfigure[]{%
  \includegraphics[width=\textwidth]{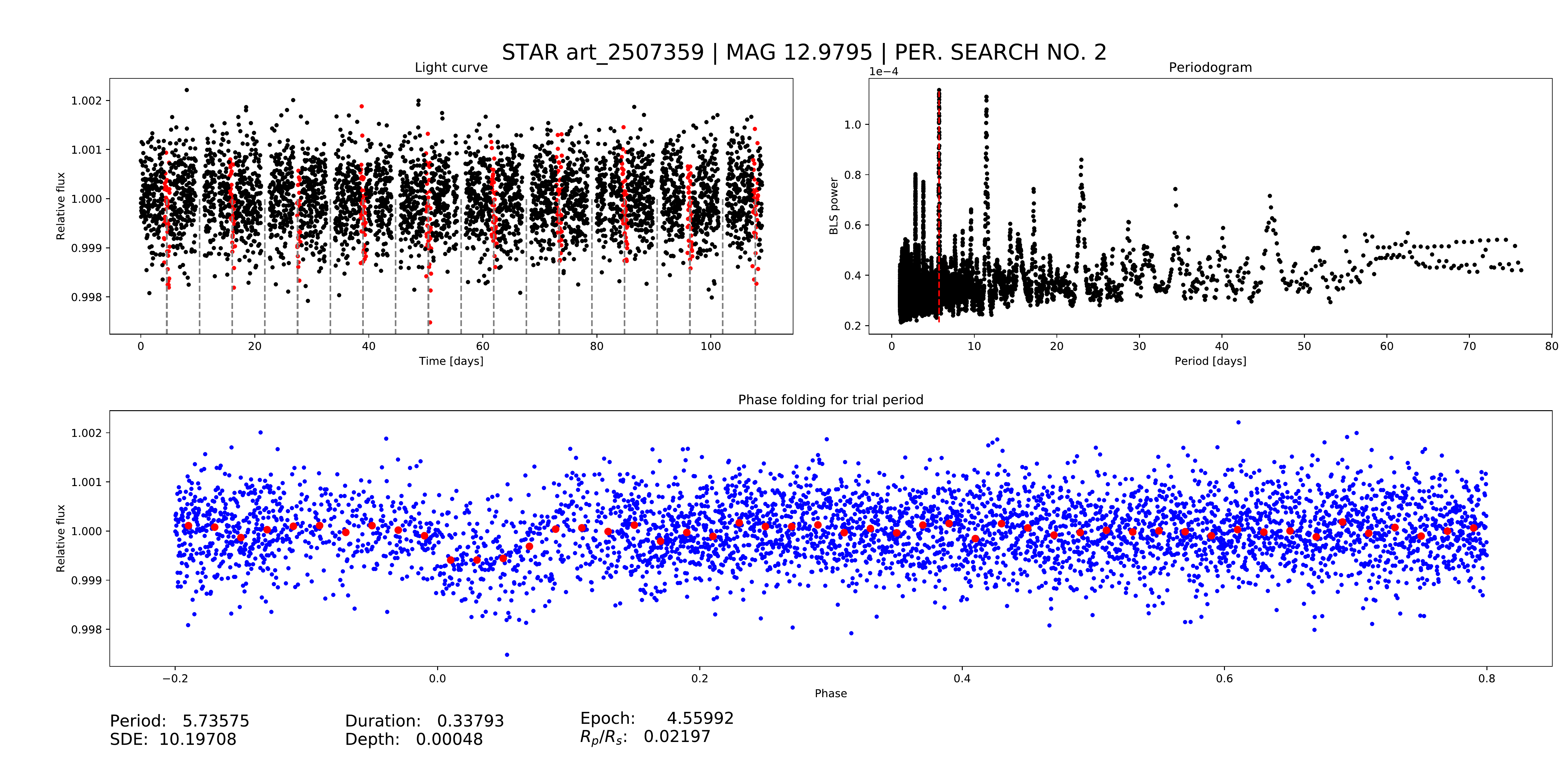}}
  \caption{\small Pipeline output for a $V=13.0$ star observed for 109.6 days (or 4 TESS sectors). A primary (at phase 0) and a secondary (at phase 0.5) were injected into the light curve. Panels refer to the first (a) and second (b) period search. Notice that all primary transits have been masked out from the light curve prior to the second period search.}\label{fig:secondary}
\end{figure*}

We test for the presence of a secondary transit/eclipse by (i) looking for two closely spaced periods (at most 0.5 days apart) and (ii) assessing whether the corresponding depths differ by at least 10\,\% (relative to the larger of the two depths). The former step also involves checking if a newly detected period matches the 1st overtone of the primary (i.e., $P_{\rm out}^{\rm primary}/2$). Figure \ref{fig:secondary} illustrates this procedure. The light curve has nine injected primary transits. These are easily recovered during the first period search despite the clear presence of a secondary at phase 0.5 (see top panel). A planet-to-star flux ratio of $5\times10^{-4}$ was assumed when simulating the secondary. Such an arbitrarily large planet-to-star flux ratio -- as a term of comparison, the occultation depth of Kepler-91b is $\sim50\:{\rm ppm}$ -- is employed here for illustrative purposes only and no attention has been paid to determine whether or not it is physically sound. When performing the second period search (after masking out all primary transits from the light curve), we notice how the BLS power associated with the true period is brought down relative to that of its 1st overtone (see bottom panel). Detection of the 1st overtone of the primary (after its removal) is a telltale sign of the presence of a secondary transit/eclipse.

\section{Candidate vetting}\label{sec:vetting}
We will be using TESS photometry alone in a first attempt to separate the expected transit-like signals from (the notionally many) astrophysical false positives (due, e.g., to EBs and BEBs) and systematic false alarms. A clear distinction will not always be possible, of course, but this will aid retaining those cases we think are due to genuine transits of short-period gas giants around LLRGB stars.

Housekeeping operations such as retaining only those systems with two or more detected transits as well as measured positive depths will be implemented. An \textit{SDE} threshold will be adopted, its value depending on the number of TESS sectors (cf.~Sect.~\ref{sec:fp}). Any measured transit depth in excess of 1\,\% will be attributed to an eclipsing binary (a $2\,R_{\rm J}$ planet transiting an LLRGB will at most cause a $\sim$0.5\,\% flux reduction). As illustrated in the previous section, the pipeline tests for a number of additional features, namely, the presence of a secondary transit/eclipse and sinusoidal behavior. Although not decisive with respect to the vetting procedure, these flags provide useful information that can be used during the detailed transit fitting. We will pay particular attention, at low \textit{SDE}, to possible period misidentifications (due to an out-of-bounds true period or else statistical in nature) by carefully inspecting the light curve and associated BLS periodogram. We also advocate for the independent fitting -- as part of the detailed transit fitting -- of the phase-folded odd and even primary transits, since a significant difference in their depths may indicate we are dealing with an eclipsing binary.

\section*{Acknowledgments}
The project leading to this publication has received funding from the European Union's Horizon 2020 research and innovation programme under the Marie Sk\l{}odowska-Curie grant agreement No.~792848 (PULSATION). This work was supported by FCT -- Funda\c{c}\~{a}o para a Ci\^{e}ncia e a Tecnologia through national funds and by FEDER through COMPETE2020 -- Programa Operacional Competitividade e Internacionaliza\c{c}\~{a}o by these grants: UID/FIS/04434/2013 \& POCI-01-0145-FEDER-007672 and PTDC/FIS-AST/28953/2017 \& POCI-01-0145-FEDER-028953. S.C.C.B.~acknowledges Investigador FCT contract IF/01312/2014/CP1215/CT0004. The research leading to the presented results has received funding from the European Research Council under the European Community's Seventh Framework Programme (FP7/2007-2013) / ERC grant agreement no.~338251 (StellarAges).

\bibliographystyle{phostproc}
\bibliography{draft.bib}

\end{document}